# Temporal Signature of Bosonic Stimulation Induced by Dark Exciton in a Two-photon Pumped Polariton Condensate

Nadav Landau[1,*], Dmitry Panna[1], Sarit Feldman[1], Sebastian Brodbeck[2], Christian Schneider[3], Sven Höfling[2], and Alex Hayat[1]

[1]*Department of Electrical and Computer Engineering, Technion – Israel Institute of Technology, Haifa, 3200003, Israel*
[2]*Technische Physik, Universität Würzburg, Am Hubland, D-97074 Würzburg, Germany*
[3]*Institute of Physics, Carl von Ossietzky Universität Oldenburg, D-26111 Oldenburg, Germany*

Strongly-coupled light-matter exciton-polaritons constitute an on-chip solid-state platform where the macroscopic quantum phenomenon of condensation is not only achievable at elevated temperatures, but can also be optically controlled, including through their interaction with inaccessible "dark" states. A recent study has shown that polariton condensation can be established under nonlinear two-photon pumping, paving the way towards dark state-condensate coherent control and highly-efficient terahertz (THz) lasing. In this letter, we show for the first time a temporal signature of dark exciton induced bosonic stimulation by investigating one- and two-photon pumped condensation with time-resolved photoluminescence spectroscopy. Our results show a clear difference in the detuning dependence of the buildup and relaxation rates of the condensate-induced blueshifts under one- and two-photon pumping. This difference is associated with the stronger exciton-fraction dependence expected for one-photon pumped condensation, where polariton-polariton stimulation dominates, compared with two-photon pumped condensation, where a 2p-exciton-to-lower-polariton THz transition can provide another stimulation channel, together with various spin-flip, electron-hole exchange and phonon-based mechanisms. These observations indicate the presence of a new dark state originated stimulation channel that could facilitate highly-efficient THz lasing in semiconductor microcavities.

Quantum condensates have been around ever since the century-old predictions of Einstein [1] and Bose [2]. Such condensates were observed in the laboratory with cold atoms [3,4] and superfluids [5], with photons in masers [6] and lasers [7], and with electron pairs in superconductors [8]. Today, they stand at the center of the quantum technology revolution, from their use in precision measurements [9] to quantum computing [10]. In semiconductor microcavities, confined light can strongly-couple to material excitations, forming composite bosonic quasiparticles known as exciton-polaritons [11]. These quasiparticles have been extensively researched over the past several decades, partly because they allow for on-chip room-temperature bosonic condensation, paving the way towards practical quantum technologies. Such condensation has been shown under both optical [12,13] and electrical [14] excitation schemes, and in both quasi- and nonequilibrium [15,16] conditions.

A particular interest exists in probing or interacting these condensates with "dark" exciton states in the cavity-embedded Quantum Wells (QWs) [17-19], inaccessible through usual one-photon excitation schemes. Such metastable states serve as important building blocks in solid-state quantum memory applications [20], owing to their long lifetimes and coherence, and in this context could be used to imprint potential landscapes on these condensates or coherently control them. Optical selection rules around $k_{\parallel} \cong 0$ dictate they can be accessed via two-photon absorption, a nonlinear optical process where two pump photons are simultaneously absorbed in the material to generate one exciton. Theoretical studies have shown that two-photon pumping the QW 2p dark exciton can lead to stimulated Terahertz (THz) emission when feeding a polariton condensate [21,22]. This opens the door to the development of a new highly-efficient THz laser source, addressing a long-standing challenge in the scientific community [23]. While the microscopic temporal dynamics at play in a polariton condensation process, typically generated by nonresonant one-photon excitation, are now well-established following rigorous theoretical works [24-26], corroborated by time-resolved experiments [27-29], the nonlinear optical excitation of these condensates via dark states remains unexplored. Two-photon generation of uncondensed polaritons has been shown [30-35], and the 2p-to-1s-exciton-based transition was also investigated using polariton bistability [19] and THz time-domain spectroscopy [17,36]. Nevertheless, polariton condensation via two-photon absorption was only very recently observed, yet still without any time-resolved information [37], showing no evident trace of dark exciton induced stimulation.

Here, we observe for the first time a temporal signature of dark exciton induced bosonic stimulation in a two-photon pumped polariton condensate, by performing a time-resolved Photoluminescence (PL) study of its dynamics. This is achieved by collecting the PL emanating from the microcavity sample following both one- and two-photon nonresonant pulsed excitation into a spectrometer-coupled Streak camera, as shown in Fig. 1 (b), and scanning the excitation power in both cases from the sub- to the above-threshold regimes, for various 1s exciton-cavity detuning energies $\Delta$. The resulting spectrally- and temporally-resolved emission pulses around the Lower Polariton (LP) resonance are shortened in time upon crossing the condensation threshold, and are subsequently extracted above that threshold for both one- and two-photon pumping at a roughly fixed value of the LP integrated PL. Though these correspond to slightly different excitation powers above threshold due to sample nonuniformity, this approach facilitates the comparison between similar LP densities without knowledge of the internal injection efficiencies in each case. Our findings show a clear difference between the one- and two-photon pumping regimes, in the detuning dependence of the buildup and relaxation rates of the condensate-induced blueshift. In the case of one-photon pumped condensation, the buildup rate of the condensate-induced blueshift grows more strongly with increasing exciton-fraction of the LPs compared with two-photon pumping, while the relaxation rate of the condensate-induced blueshift follows a similar trend. This indicates that in the two-photon pumping regime, polariton-polariton scattering along the LP branch, which depends strongly on the LP exciton fraction, is no longer the sole contributor to the bosonic final state stimulation of the LP ground state. Rather, the 2p-to-1s-exciton-based LP THz transition most likely comes into play [21,22], as it depends less on the LP exciton fraction. We corroborate this by qualitatively fitting our data to a semi-classical Boltzmann rate equations model that captures the dynamics using phenomenological constants and a minimal number of free parameters, showing good agreement. The deduced presence of another bosonic stimulation channel to the usual LP-LP scattering in the two-photon pumped condensation, more weakly dependent on the LP exciton fraction, is a crucial telltale sign and potential steppingstone towards polariton-based doubly-stimulated THz emission and solid-state coherent control of macroscopic quantum states using dark states.

*Contact author: nlandau@campus.technion.ac.il

Our sample consists of a λ/2 AlAs cavity layer embedded between two distributed Bragg reflectors (DBRs) comprised of 23 (27) alternating layers of $Al_{0.2}Ga_{0.8}As$ and AlAs in the lower (upper) DBR, respectively, giving a cavity Q-factor of ~5000 and cavity mode lifetime of $\tau_{cav} \sim 2\mathrm{ps}$. Three stacks of four ~7nm GaAs QWs with ~4nm AlAs barriers are located at the three antinode positions of the confined microcavity mode. The lowest QW subband 1s exciton state is measured to be at $\sim 1613\mathrm{meV}$ and the 2p "dark" exciton state is calculated from it [43] to be at $\sim 1622\mathrm{meV}$, using the experimentally extracted average QW width and characteristic literature-based GaAs parameters. The sample thickness is tapered across one axis to allow access to different 1s exciton-cavity detuning energies Δ by impinging different spots along that axis. Scanning that axis and measuring the normalized sample reflectivity spectra at each point, the usual anti-crossing indicating strong coupling can be observed (see Fig. 2 (j)), with a measured zero detuning vacuum Rabi splitting of ~13.6 meV. The sample also exhibits both one- and two-photon pumped LP condensation [37].

In our experiment, the sample was held in a closed-cycle liquid helium-flow cryostat at T~6 K. Our time-resolved one- and two-photon absorption spectroscopy was achieved by pumping the sample with the femtosecond-pulsed output of an Optical-Parametric Amplifier (OPA), operating at a repetition rate of 1 KHz. In the one-photon pumping case, our pump photon energy was tuned to the first minimum outside the microcavity stopband at the pump incidence angle (~1722meV), whereas in the two-photon pumping case, it was tuned to slightly above half the resonance energy of the QW HH1 2p dark exciton state in our sample (~821meV). Both cases facilitate highly off-resonant pulsed excitation of our system, with the latter also being resonant with half the 2p dark exciton energy. A variable neutral density filter was used to attenuate the pump beam, and a longpass filter suppressed any possible residual higher harmonics from the OPA. The pump was then focused onto the sample to a spot of area ~24000μm$^2$, at ~30° from normal incidence. The resulting PL was collected from the central area of that spot using a microscope objective of NA=0.4. It was then relayed and focused onto a spectrometer-coupled streak camera for energy- and time-resolved measurements. To simultaneously verify condensation in our sample and choose the appropriate excitation regime for both one- and two-photon pumping, the objective Fourier plane was imaged onto a second spectrometer while scanning the pump power across the condensation threshold using EMCCD-based angle-resolved PL in each case [37]. A representative sub-threshold emission image at $\Delta \cong -3.5\,\mathrm{meV}$ is shown in Fig. 2 (i) with the calculated coupled-oscillator-model dispersions [44] superimposed. Time-resolved measurements were then recorded for both excitation regimes at several 1s exciton-cavity energy detunings.

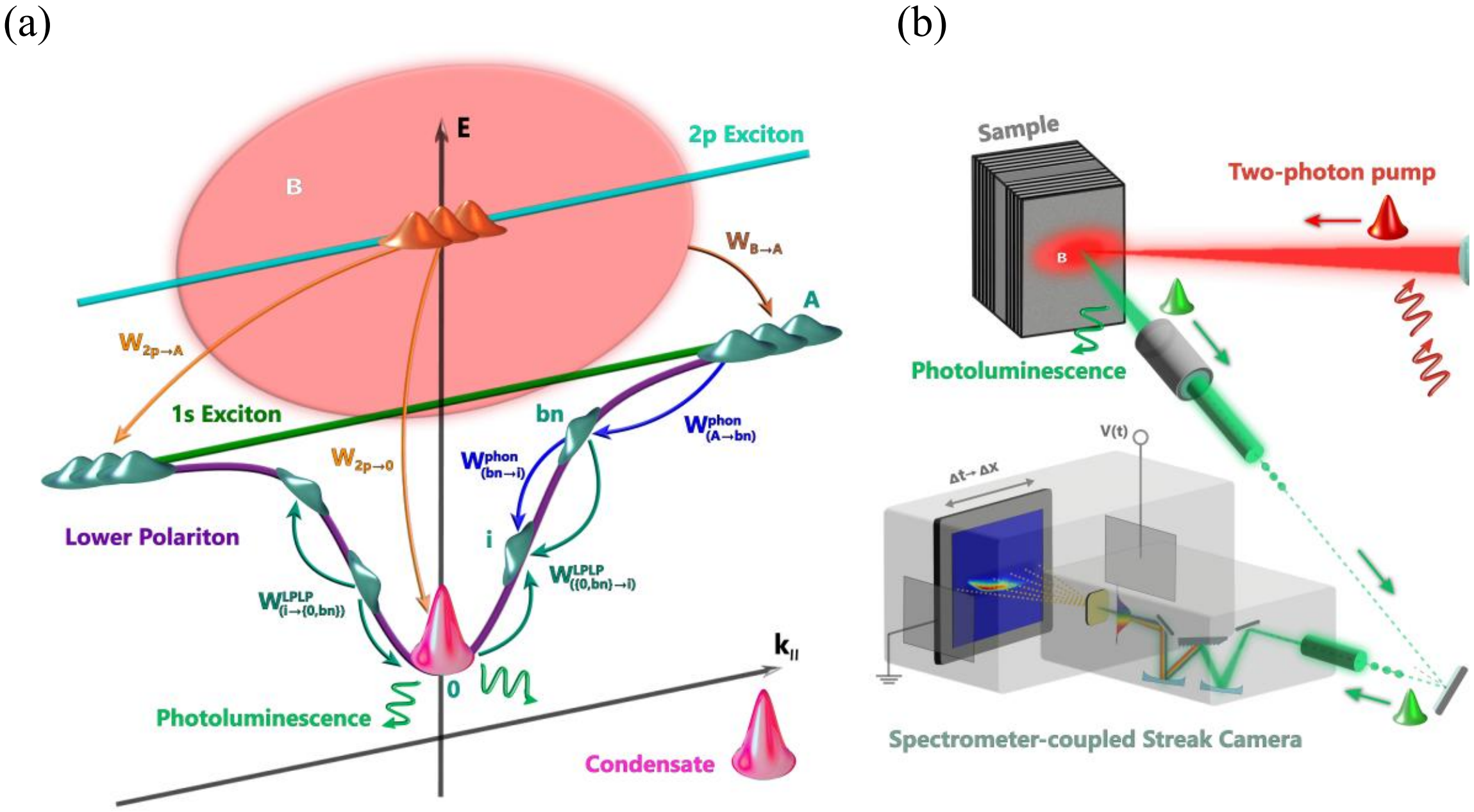

FIG. 1. Illustration of the experiment and modelled dynamics. (a) In-plane dispersions of the 2p (**teal**) and 1s (**dark green**) exciton branches together with the LP branch (**purple**), showing the one- and two-photon pumped condensation dynamics studied in our work, including the rate equation state groups and terms used in our model (see Supplemental Material [38]). A background population of excitons ($n_B$) is initially generated in the region of our one- (two-) photon pump spot (B, **red**), composed of one- (two-) photon allowed exciton states and excitons diffused from remote areas, and is schematically introduced in the band diagram. It then feeds a very high $k_{\|}$ LP reservoir $A$ at a rate $W_{B\to A}$ via phonon, electron-hole exchange and spin-flip mechanisms, after which phonon-based and LP-LP relaxation takes place. For two-photon pumping, direct excitation of 2p excitons can also occur, followed by either THz feeding of the ground state ($W_{2p\to 0}$) or phonon-based feeding of reservoir $A$ ($W_{2p\to A}$). (b) The two-photon pump (**red**) is focused onto the microcavity sample (top). The resulting PL (**light green**) is thereafter collected onto a spectrometer-coupled streak camera for energy- and time-resolved spectroscopy (bottom).

As can be seen in Fig. 2 (a)-(h), PL from the LP line at all detunings in the experiment and for both one- and two-photon pumping above threshold exhibits two evident time-dependent blueshifts - one larger blueshift, present immediately at the beginning of the emission pulse, that diminishes much slower with time, and another smaller but sharper blueshift appearing at a certain detuning-dependent delay time after the beginning of the pulse. This latter spectral bump reduces much quicker for the remainder of the pulse duration. We associate the first blueshift with the large background exciton population generated in the region of our pump spot [29], which shifts the LP energy upwards and then gradually lowers it as that population is depleted. The second blueshift is an interaction-induced condensate blueshift that inherently depends on the condensate population, and which takes some time to build up. Our measurements indicate a clear, detuning-dependent difference in this buildup between one- and two-photon pumping. A similar pattern is observed for the condensate-induced blueshift relaxation - it depends more strongly on the LP exciton fraction for the case of one-photon pumping, and is also faster at all detunings. We attribute these differences to the fact that in one-photon pumped condensation, these rates are expected to have stronger dependence on the LP exciton-fraction, since the bosonic final-state stimulation is governed by polariton-polariton scattering along the LP branch. In contrast, in the two-photon pumped condensation, the stimulation process is expected to contain some contribution from the 2p-to-LP THz transition which has a weaker LP exciton fraction dependence, making the overall dependence weaker. We stress here that while two-photon excitation provides more stimulation channels for the condensate, these are of lower transition probability, leading to one-photon excitation populating the condensate more efficiently for the same integrated PL value (See Fig. 2 (a)-(h)).

To theoretically model the relaxation dynamics in our experiment and because the present study analyses only population-driven, time-resolved PL intensities and energy blueshifts that are not explicitly affected by coherence, we have implemented a semi-classical Boltzmann rate equation model with phenomenological constants and a minimal number of free parameters to obtain a qualitative fit to the data (see Supplemental Material [38]). A full density-matrix approach

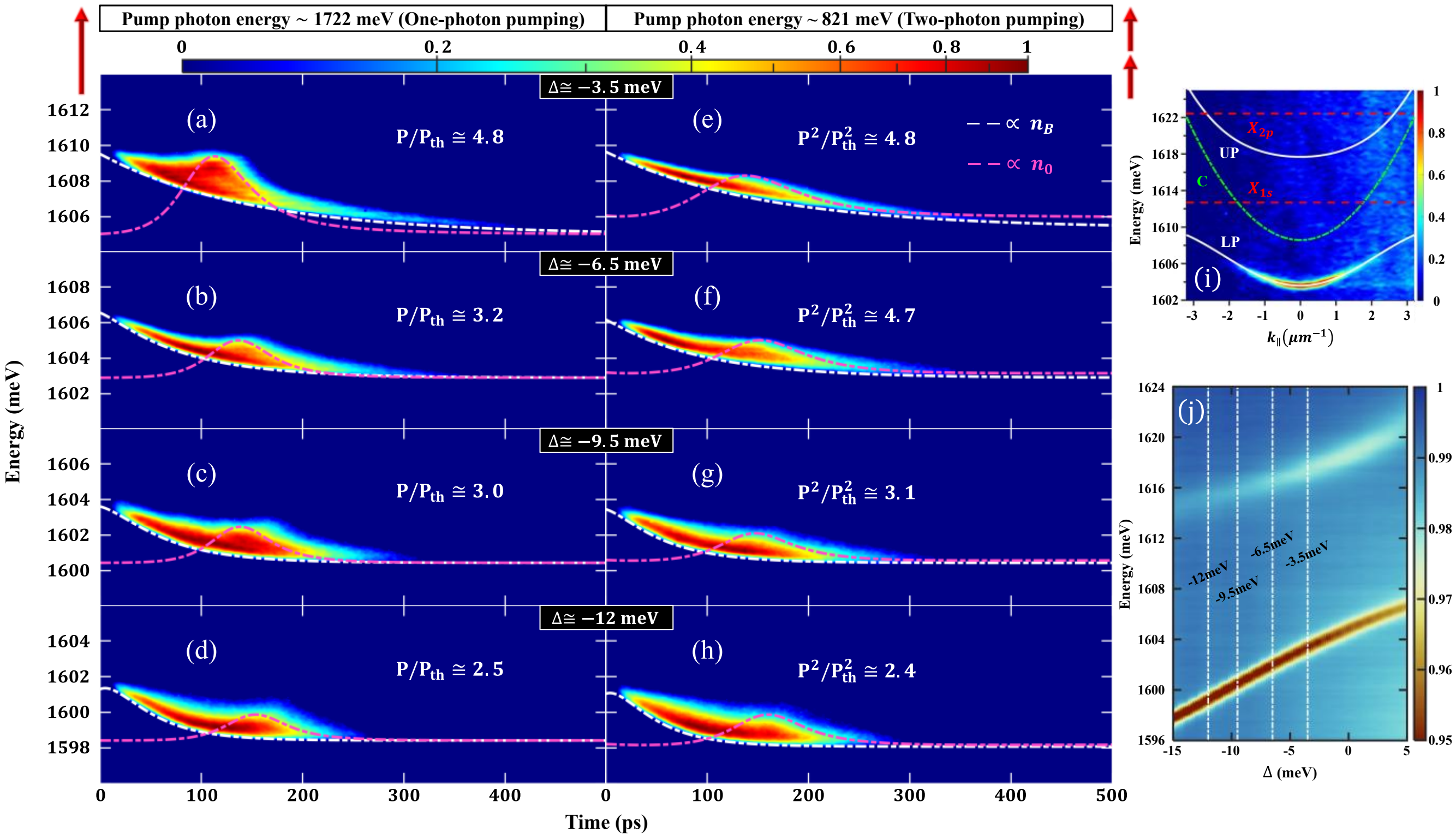

FIG. 2. Time-resolved LP condensate PL for several 1s exciton-cavity detuning energies $\Delta$ and at $T \cong 6\mathrm{K}$, for both one- (a)-(d) and two-photon (e)-(h) pumping. The PL (color) in each panel is normalized and log-scaled for better visibility, with the color code indicated on top. All cases were extracted for a similar integrated PL value above the condensation threshold, with the resulting one- and two-photon threshold-normalized excitation powers shown on each plot. In all cases, the emission is initially blueshifted by several meV, depending on the LP exciton fraction, and then reduces back down on a timescale of several hundred ps. This is associated with the background exciton population in our pump spot region $n_B$ (dashed-white). Thereafter, a second, smaller but sharper blueshift occurs before the PL signal dies out, associated with the LP condensate population $n_0$ (**dashed-pink**). Buildup and relaxation rates of this latter blueshift vary not only with the LP exciton fraction but between one- and two-photon pumping (stronger dependence for the former). (i) A representative sub-threshold angle-resolved PL image at $\Delta \cong -3.5\,\mathrm{meV}$ with the calculated dispersions of the polaritons (white), 1s ($X_{1s}$) and 2p ($X_{2p}$) excitons (dashed red) and cavity mode (C, dashed green) superimposed. (j) Measured normalized sample reflectivity (log scale) at $T \cong 4\mathrm{K}$ as a function of energy and detuning, showing the typical anti-crossing indicating strong coupling. The four detunings in (a)-(h) are marked in white dashed vertical lines and black text.

[45] was not pursued here since no phase-resolved observables were measured. Similar models have been very successful in capturing the relaxation dynamics of microcavity polaritons and their condensation without explicitly invoking coherence [24-26,28,29], and we employ a similar one to Ref. [29], but with more equations describing relaxation along the LP branch as well as the 2p dark exciton population in the case of two-photon pumping. The extension is warranted in order to properly capture the important experimentally observed one- and two-photon pumped condensation dynamics and in particular to compare between the different final-state stimulation mechanisms. Following similar arguments, we denote the populations and average number of states in any state group $G$ in our model as $n_G = \sum_{k_\parallel \in \{G\}} n_{k_\parallel}$ and $N_G = \sum_{k_\parallel \in \{G\}} 1$, respectively, where $n_{k_\parallel}$ denotes the population at a particular $k_\parallel$ state. These state groups and their dynamics (including pumping and decay terms $P_G, \gamma_{P_G}, \gamma_G$, as well as scattering rates $W_{G_i \to G_j}$) are shown in Fig. 1 (a) and the equations follow the typical form for bosonic quasiparticles [29,44]:

$$\dot{n}_G = P_G e^{-\gamma_{P_G} t} - \gamma_G n_G - W_{G \to G_i} n_G \left(\frac{n_{G_i}}{N_{G_i}} + 1\right) + W_{G_j \to G} n_{G_j} \left(\frac{n_G}{N_G} + 1\right)$$
$$G = 2p, B, A, bn, i, 0 \qquad (1)$$

with the pumping term chosen as a decaying exponent in time so as to keep it in general form [29]. The different state groups are described in detail in the Supplemental Material [38], where the full set of equations for both one- and two-photon pumping is explicitly presented. Of these we note first the background exciton population formed within our pump spot at early times following nonresonant one- or two-photon pulsed excitation $n_B$, which we attribute to the large and slowly diminishing initial-time blueshifts of our LP emission pulses in Fig. 2 (a)-(h). And second, the 2p "dark" exciton population $n_{2p}$, which is only present for two-photon excitation. Theirs are the only two equations where a pumping term is included (first term on the right-hand side of Eqn. (1)). Applying the model to our data, with the pumping terms acting as free fit parameters and literature-based values (adjusted to our extracted LP dispersions) introduced for the other terms (see Supplemental Material [38] for a full classification), a qualitative fit can be achieved (dashed-white lines in Fig. 2 (a)-(h) are proportional to $n_B(t)$, while dashed-pink lines are proportional to $n_0(t)$). This proportionality stems from the linear relations between the background exciton and condensate populations and the magnitudes of their induced LP blueshifts. From the parameter choice enabling a good fit to our data, we deduce that for the presence of the THz-assisted ($W_{2p \to 0}$) stimulation channel to have any visible influence on our observed two-photon pumped dynamics, especially for the condensate-induced blueshift relaxation and its detuning dependence, the 2p exciton pumping term $P_{2p}$ must be at least an order-of-magnitude larger than the background exciton pumping term $P_B$. This can be seen in Fig. 2 (e)-(h), giving a better fit to our model, yet we were unable to reach a regime where this channel dominates the dynamics, making it difficult to determine its weight in them.

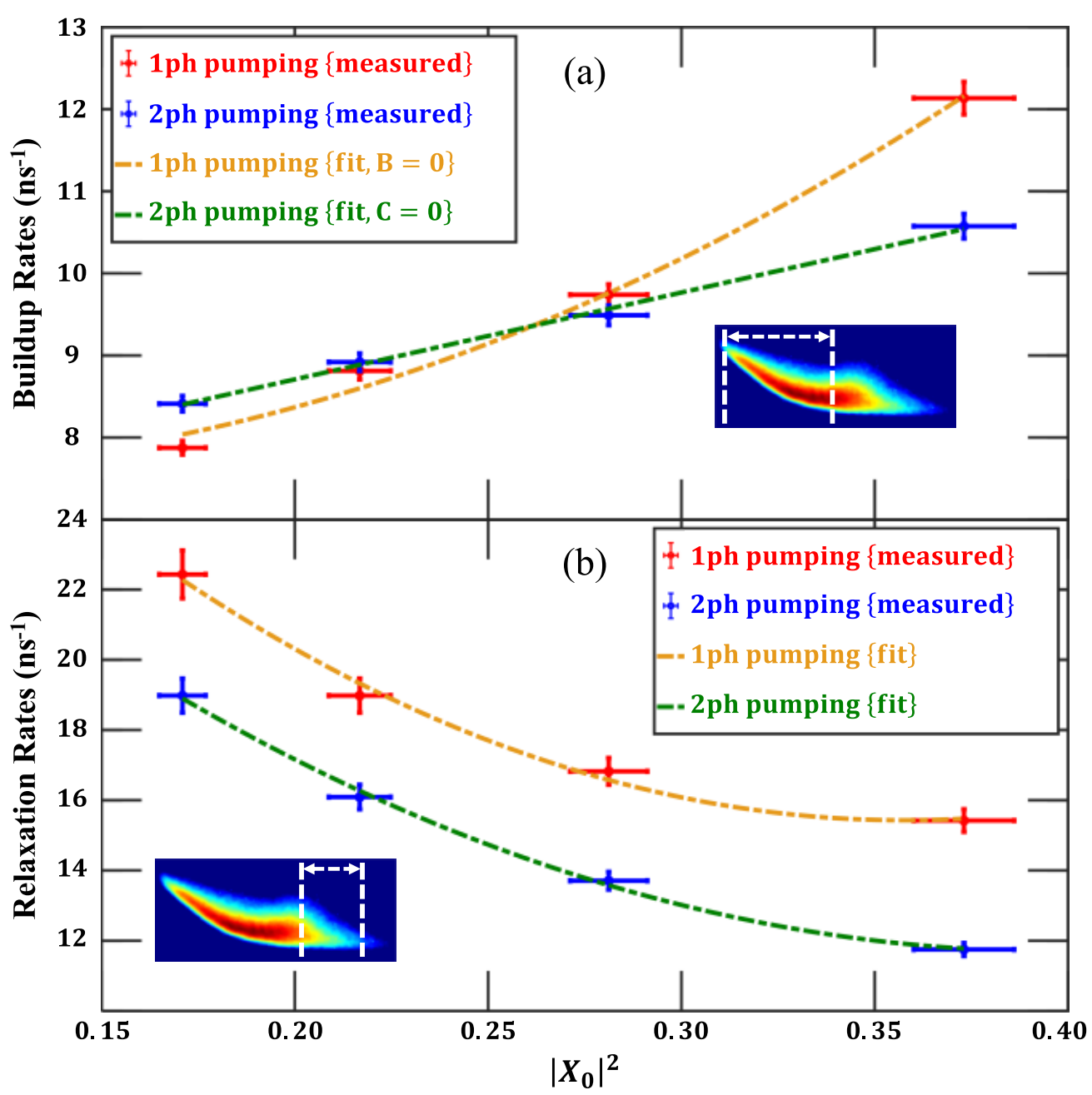

FIG. 3. Condensate-induced blueshift buildup (a) and relaxation (b) rates (in $\mathrm{ns}^{-1}$) as a function of the LP ground state exciton fraction $|X_0|^2$. Rates are extracted from the data for both one- (**red**) and two-photon (**blue**) pumping at all four detunings in the experiment, and fitted to $y = A(1-x) + Bx + Cx^2$ with $x = |X_0|^2$ (**dashed-gold** and **dashed-green** lines, respectively). Fits constraining B and C to zero for the one- and two-photon pumped buildup rates, respectively, reflect the assumed dominant stimulation channel in each case and confirm the expected trend, which is clearly stronger for one-photon pumping. For the relaxation rates, the dynamics are less trivial, and all three fitting parameters were required to yield robust fits to the data. The insets in panels (a) and (b) visualize, respectively, the extraction of the buildup and relaxation rates shown in the plots. A more detailed description of the fits is given in the Supplemental Material [38].

To substantiate the presence of this stimulation channel, we show in Fig. 3 the extracted buildup and relaxation rates of the condensate-induced blueshifts for both one- and two-photon pumping, as functions of the LP ground state exciton fraction $|X_0|^2$. The plots were qualitatively fitted to $y = A(1-x) + Bx + Cx^2$ $(x = |X_0|^2)$, where A, B and C act as fit parameters. The first term $A(1-x)$ relates to the condensate radiative decay rate but also includes any blueshift baseline present in the data. The second (linear in-$|X_0|^2$) term $Bx$ represents any single LP contributions to the dynamics, such as residual reservoir interactions and most importantly the possible 2p-to-LP THz stimulation under two-photon pumping. The third (quadratic in-$|X_0|^2$) term $Cx^2$ represents any LP-LP-based contributions, such as the LP-LP stimulation under both pumping regimes as well as LP-LP-based out-coupling from the condensate. The data shows a clear difference in the dependence between one- and two-photon pumping, as can already be seen in Fig. 2 (a)-(h). We compare unconstrained fits (containing A, B and C) to fits constraining $B$ and $C$ to 0 for the one- and two-photon pumped data, respectively, reflecting the assumed dominant stimulation channel in each case.

All cases yield very tight fits to our data except for the constrained fits to the blueshift relaxation rates, where both linear and quadratic terms in $|X_0|^2$ are required, and constraining either B or C to zero strongly degrades the fit quality (see Supplemental Material [38] for a more detailed analysis). For the unconstrained fit to the buildup rate under one-photon pumping, $C \cong 4.5B$ and a very tight fit remains even when constraining $B$ to zero (Fig. 3 (a), dashed-gold), so we deduce that LP-LP-based stimulation dominates these dynamics, as expected. For the unconstrained fit to the buildup rate under two-photon pumping, $B \cong 2C$ and a very tight fit remains even when constraining $C$ to zero (Fig. 3 (a), dashed-green), so we deduce that the 2p-to-LP THz stimulation is likely present in these dynamics, as expected. For the relaxation rates under both pumping regimes, our fitting implies that this stage of the dynamics is dominated by a competition between the condensate's radiative decay and LP-LP-based out-coupling and its reservoir-based replenishment, with the stimulated replenishment being less pronounced so that no similar distinction can be made.

Altogether, the dependence of the condensate-induced blueshift dynamics on $|X_0|^2$ clearly differs between one- and two-photon pumping and is moreover consistent with the expected dependence of the two distinct stimulation channels in each case. These findings indicate that LP-LP scattering is no longer the sole contributor to the bosonic stimulation in the case of two-photon pumped condensation, and together with the qualitative fit to our model, we deduce the presence of another stimulation channel with weaker $|X_0|^2$ dependence, likely the 2p-to-LP THz transition [21].

In conclusion, we have shown a temporal signature of dark exciton induced bosonic stimulation in the time-resolved measurement of a two-photon pumped polariton condensate. Our findings, achieved in a GaAs-based microcavity, indicate a clear difference between the usual one-photon pumping scheme and the new two-photon pumping scheme, in the detuning dependence of the condensate-induced blueshift's buildup and relaxation rates. A semi-classical Boltzmann rate equation model encompassing the investigated processes shows a qualitative fit to the experimental data and, along with the extracted rates, points to the presence of a stimulation channel other than the usual LP-LP scattering, most likely that of the 2p dark exciton originated THz transition suggested in [21]. This finding could lead to the development of polariton-based doubly-stimulated THz lasing. Future work should measure time-resolved THz amplification in the sample directly [17,36] when two-photon pumping the system above the condensation threshold.

Acknowledgments: The authors acknowledge the financial support of the Israel Science Foundation (ISF Grants No. 934/18 and No. 1605/22) and the German Research Foundation (DFG) within the projects HO5194/12-1 and HO 5194/20-1.

Data availability: The data that support the findings of this study are openly available in Zenodo [42].

[1] A. Einstein, Quantum theory of the monatomic ideal gas, Sitzungsber. Preuss. Akad. Wiss., *Phys.-Math. Kl.* **261**, 44 (1924).

[2] S. N. Bose, Plancks gesetz und lichtquantenhypothese, *Z. Phys.* **26**, 178 (1924).

[3] M. H. Anderson, J. R. Ensher, M. R. Matthews, C. E. Wieman, and E. A. Cornell, Observation of Bose-Einstein condensation in a dilute atomic vapor, *Science* **269**, 198 (1995).

[4] K. B. Davis, M. O. Mewes, M. R. Andrews, N. J. van Druten, D. S. Durfee, D. M. Kurn, and W. Ketterle, Bose-Einstein condensation in a gas of sodium atoms, *Phys. Rev. Lett.* **75**, 3969 (1995).

[5] P. Kapitza, Viscosity of liquid helium below the λ-point, *Nature* **141**, 74 (1938) ; J. F. Allen and A. D. Misener, Flow of liquid helium II, *Nature* **141,** 75 (1938).

[6] J. P. Gordon, H. J. Zeiger, and C. H. Townes, The Maser - new type of microwave amplifier, frequency standard, and spectrometer, *Phys. Rev.* **99**, 1264 (1955).

[7] A. L. Schawlow and C. H. Townes, Infrared and optical masers, *Phys. Rev.* **112**, 1940 (1958).

[8] J. Bardeen, L. N. Cooper and J. R. Schrieffer, Theory of superconductivity, *Phys. Rev.* **108**, 1175 (1957).

[9] D. Becker, M. D. Lachmann, S. T. Seidel, *et al.,* Space-borne Bose–Einstein condensation for precision interferometry, *Nature* **562**, 391 (2018).

[10] F. Arute, K. Arya, R. Babbush, *et al.,* Quantum supremacy using a programmable superconducting processor, *Nature* **574**, 505 (2019).

[11] C. Weisbuch, M. Nishioka, A. Ishikawa, and Y. Arakawa, Observation of the coupled exciton-photon mode splitting in a semiconductor quantum microcavity, *Phys. Rev. Lett.* **69**, 3314 (1992).

[12] H. Deng, G. Weihs, D. Snoke, J. Bloch, and Y. Yamamoto, Polariton lasing vs. photon lasing in a semiconductor microcavity, *Proc. Natl. Acad. Sci. U.S.A.* **100**, 15318 (2003).

[13] J. Kasprzak, M. Richard, S. Kundermann, *et al.,* Bose–Einstein condensation of exciton polaritons, *Nature* **443**, 409 (2006).

[14] C. Schneider, A. Rahimi-Iman, N. Y. Kim, *et al.,* An electrically pumped polariton laser, *Nature* **497**, 348 (2013).

[15] M. Richard, J. Kasprzak, R. André, R. Romestain, L. S. Dang, G. Malpuech, and A. Kavokin, Experimental evidence for nonequilibrium Bose condensation of exciton polaritons, *Phys. Rev. B* **72**, 201301 (2005).

[16] M. Wouters, I. Carusotto and C. Ciuti, Spatial and spectral shape of inhomogeneous nonequilibrium exciton-polariton condensates, *Phys. Rev. B* **77**, 115340 (2008).

[17] J. M. Ménard, C. Pöllmann, M. Porer, U. Leierseder, E. Galopin, A. Lemaître, A. Amo, J. Bloch, and R. Huber, Revealing the dark side of a bright exciton–polariton condensate, *Nat. Commun.* **5**, 4648 (2014).

[18] D. Schmidt, B. Berger, M. Kahlert, M. Bayer, C. Schneider, S. Höfling, E. S. Sedov, A. V. Kavokin, and M. Aßmann, Tracking dark excitons with exciton polaritons in semiconductor microcavities, *Phys. Rev. Lett.* **122**, 047403 (2019).

[19] E. Rozas, E. Sedov, Y. Brune, S. Höfling, A. Kavokin, and M. Aßmann, Polariton–dark exciton interactions in bistable semiconductor microcavities, *Phys. Rev. B* **108**, 165411 (2023).

[20] E. Poem, Y. Kodriano, C. Tradonsky, N. H. Lindner, B. D. Gerardot, P. M. Petroff, and D. Gershoni, Accessing the dark exciton with light, *Nat. Phys.* **6**, 993 (2010).

[21] A. V. Kavokin, I. A. Shelykh, T. Taylor, and M. M. Glazov, Vertical cavity surface emitting terahertz laser, *Phys. Rev. Lett.* **108**, 197401 (2012).

[22] G. Slavcheva and A. V. Kavokin, Polarization selection rules in exciton-based terahertz lasers, *Phys. Rev. B* **88**, 085321 (2013).

[23] G. Davies and E. Linfield, Bridging the terahertz gap, Phys. World **17**, 37 (2004).

[24] F. Tassone and Y. Yamamoto, Exciton-exciton scattering dynamics in a semiconductor microcavity and stimulated scattering into polaritons, *Phys. Rev. B* **59**, 10830 (1999).

[25] D. Porras, C. Ciuti, J. J. Baumberg, and C. Tejedor, Polariton dynamics and Bose-Einstein condensation in semiconductor microcavities, *Phys. Rev. B* **66**, 085304 (2002).

[26] T. D. Doan, H. T. Cao, D. B. Tran Thoai, and H. Haug, Condensation kinetics of microcavity polaritons with scattering by phonons and polaritons, *Phys. Rev. B* **72**, 085301 (2005).

[27] H. Deng, G. Weihs, C. Santori, J. Bloch, and Y. Yamamoto, Condensation of semiconductor microcavity exciton polaritons, *Science* **298**, 199 (2002).

[28] J. Kasprzak, D. D. Solnyshkov, R. André, L. S. Dang, and G. Malpuech, Formation of an exciton polariton condensate: thermodynamic versus kinetic regimes, *Phys. Rev. Lett.* **101**, 146404 (2008).

[29] M. De Giorgi, D. Ballarini, P. Cazzato, *et al.,* Relaxation oscillations in the formation of a polariton condensate, *Phys. Rev. Lett.* **112**, 113602 (2014).

[30] G. Leménager, F. Pisanello, J. Bloch, *et al.,* Two-photon injection of polaritons in semiconductor microstructures, *Opt. Lett.* **39**, 307 (2014).

[31] J. Schmutzler, M. Aßmann, T. Czerniuk, M. Kamp, C. Schneider, S. Höfling, and M. Bayer, Nonlinear spectroscopy of exciton-polaritons in a GaAs-based microcavity, *Phys. Rev. B* **90**, 075103 (2014).

[32] N. Lundt, Ł. Dusanowski, E. Sedov, *et al.,* Optical valley Hall effect for highly valley-coherent exciton-polaritons in an atomically thin semiconductor, *Nat. Nanotechnol.* **14**, 770 (2019).

[33] X. Liu, J. Yi, Q. Li, S. Yang, W. Bao, C. Ropp, S. Lan, Y. Wang, and X. Zhang, Nonlinear optics at excited states of exciton polaritons in two-dimensional atomic crystals, *Nano Lett.* **20**, 1676 (2020).

[34] M. Steger, C. Gautham, D. W. Snoke, L. Pfeiffer, and K. West, Slow reflection and two-photon generation of microcavity exciton–polaritons, *Optica* **2**, 1 (2015).

[35] C. Gautham, M. Steger, D. Snoke, K. West, and L. Pfeiffer, Time-resolved two-photon excitation of long-lifetime polaritons, *Optica* **4**, 118 (2017).

[36] J. L. Tomaino, A. D. Jameson, Y. S. Lee, G. Khitrova, H. M. Gibbs, A. C. Klettke, M. Kira, and S. W. Koch, Terahertz excitation of a coherent

Λ-type three-level system of exciton-polariton modes in a quantum-well microcavity, *Phys. Rev. Lett.* **108**, 267402 (2012).
[37] N. Landau, D. Panna, S. Brodbeck, C. Schneider, S. Höfling, and A. Hayat, Two-photon pumped exciton-polariton condensation, *Optica* **9**, 1347 (2022).
[38] See Supplemental Material for a detailed description of the rate equation model and its terms and parameters, a linear-scaled version of the experimental data, a similar log-scaled version for a lower IPL value above threshold and extracted input-output curves at all investigated detunings, which includes Refs. [39–42].
[39] P. G. Savvidis, J. J. Baumberg, R. M. Stevenson, M. S. Skolnick, D. M. Whittaker, and J. S. Roberts, Angle-resonant stimulated polariton amplifier, *Phys. Rev. Lett.* **84**, 1547 (2000).
[40] H. Deng, H. Haug, and Y. Yamamoto, Exciton-polariton Bose-Einstein condensation, *Rev. Mod. Phys.* **82**, 1489 (2010).
[41] S. Pau, G. Björk, J. Jacobson, H. Cao, and Y. Yamamoto, Microcavity exciton-polariton splitting in the linear regime, *Phys. Rev. B* **51**, 14437 (1995).
[42] D. W. Snoke, W. W. Rahle, K. Köhler, and K. Ploog, Spin flip of excitons in GaAs quantum wells, *Phys. Rev. B* **55**, 13789 (1997).
[43] H. Mathieu, P. Lefebvre, and P. Christol, Simple analytical method for calculating exciton binding energies in semiconductor quantum wells, *Phys. Rev. B* **46**, 4092 (1992).
[44] A. Kavokin, J. J. Baumberg, G. Malpuech, and F. P. Laussy, *Microcavities* (Oxford University Press, Oxford, 2017).
[45] C. Lüders, M. Pukrop, E. Rozas, C. Schneider, S. Höfling, J. Sperling, S. Schumacher, and M. Aßmann, Quantifying quantum coherence in polariton condensates, *PRX Quantum* **2**, 030320 (2021).
[46] N. Landau, D. Panna, S. Feldman, S. Brodbeck, C. Schneider, S. Höfling, and A. Hayat, Data for: Temporal signature of bosonic stimulation induced by dark exciton in a two-photon-pumped polariton condensate (2026), Zenodo, https://doi.org/10.5281/zenodo.19520757.